# Feasible Surface Plasmon Routing Based on The Self-assembled InGaAs/GaAs Semiconductor Quantum Dot Located between Two Silver Metallic Waveguides

Myong-Chol Ko,[1] Nam-Chol Kim,[1,*] Hyok Choe,[1] Su-Ryon Ri,[1]

Chol-Won Ri,[1]

[1]Faculty of Physics, **Kim Il Sung** University, Pyongyang, DPR of Korea

**Abstract:** We proposed an experimentally feasible scheme of nano-plamonic switch and quantum router via the single self-assembled InGaAs/GaAs semiconductor quantum dot (SQD) with a V type three-level energy structure located between two silver metallic waveguides. We studied theoretically transmission and transfer rates of single plasmons in such a multi-ports system via the real-space approach, where our results showed that single plasmons from the input port could be switchable and redirected by controlling parameters, such as the intensity of the classical field, the detunings, and the interaction between the SQD and the waveguides. Our proposed scheme could be used not only in the design of quantum routers and quantum switches for the construction of quantum network, but also in quantum photonic integrated circuits.

[*] Electronic mail: nc.kim@ryongnamsan.edu.kp

## 1. Introduction

With the widely use of single photons as information carriers rather than the limited electrons, the research of the quantum information processing has rapidly growth and become one of the most promising fields [1,2]. Photons are very extensively used in quantum information processing, such as quantum network [3], quantum computation [4-7], quantum communication [8,9], and quantum devices [10-21] by controlling the state of single photons due to the speed of light and their robustness against various sources of decoherence. Generally, photons interact weakly with the external environment, therefore we need to find the good ways of controlling the quantum states of photons. However, strong coupling between single photons and quantum emitters, such as quantum dots and atomic systems, could be accomplished by constraining the photons in low dimensions such as in one dimensional (1D) nano waveguide, the transverse cross sections of which is on the order of the square of the wavelength [10].

In recent years, single plasmon switch and quantum network based on the one dimensional waveguide has become a hot topic of increasing attraction for researchers in the fields of quantum information processing, which is extensively investigated not only in theoretically [22-24], but also experimentally [25-30]. Recently, the first single photon router in the microwave regime via a superconducting transmon qubit coupled extremely efficiently to a superconducting transmission line [25] was experimentally realized and the first experimental demonstration of plasmon-exciton coupling between silver nanowire (NW) and a pair of SQD [26] was also reported, where the interparticle distance between the two SQDs range from microns to 200 nm within the diffraction limit and parameters including the SP propagation length and the wire terminal reflectivity are experimentally determined. With the rapidly development of scalable quantum control of single photons, the reseach on the quantum network which combines quantum channels with quantum nodes is taken in many systems, such as optomechanical system [9], cavity quantum electrodynamics [23,25,27], and whispering-gallery-mode resonator [31], where a fundamental element inside quantum node is just quantum router, which plays important role for controlling the quantum signal channel at a single photon level, enabling us to deliver and distribute the quantum signals to the desired node with optimal control.

Among a great deal of theoretical papers on single photon router from one channel to the other channel, most of them is based on the discrete scattering equations using the coupled cavity arrays and we could not find papers based on the real space approach, especially, a scheme of a single plasmon router with coupled SQD-plasmonic waveguide systems has not been proposed. Moreover, the experiment of photonic components based on dielectric loaded surface plasmon polariton waveguides excited by single nitrogen vacancy [32] was investigated for silver plasmonic waveguide. And it is also convenient to use the real space approach for the discussion of the single photon routing on metallic plasmonic waveguide. Motivated by the above considerations, we theoretically proposed a scheme for a single plasmon router with the single self-assembled InGaAs/GaAs SQD with a V type three-level energy structure located between two silver metallic waveguides via real space approach.

## 2. Theoretical model and dynamics equations

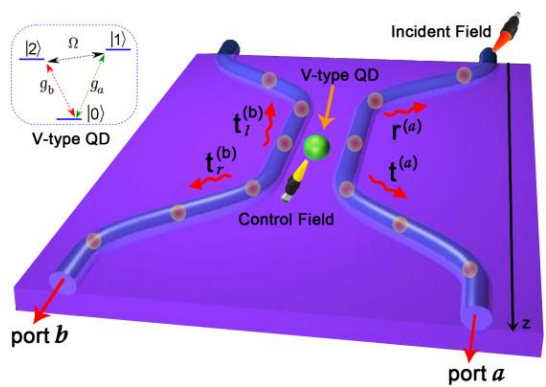

**Fig. 1** (color online). Schematic diagram of a single plasmon router composed of two silver nanowaveguides and a V type three-level SQD located between them. The energy structure of the SQD is described by a ground state, $|0\rangle$, and two excited states $|1\rangle, |2\rangle$, respectively. Two silver nanowaveguides couple to the SQD through a dipole-allowed transitions $|0\rangle \leftrightarrow |1\rangle$ and $|0\rangle \leftrightarrow |2\rangle$ with coupling constants, $g_a$ and $g_b$, respectively, and a classical field, $\Omega$, is applied to resonantly drive $|1\rangle \leftrightarrow |2\rangle$ transition. An incident single plasmon from the left side of "port $a$" could be transmitted, reflected, or transferred to "port $b$". We set the distance between the two channels as $L$.

We propose an experimentally feasible scheme of a single plasmon router, which are composed of two silver metallic waveguides and a SQD located between them as shown in Fig. 1. The SQD could be the single self-assembled InGaAs/GaAs SQD with a V type t hree-level energy structure denoted by a ground state $|0\rangle$ and two excited states, $|1\rangle$ and $|2\rangle$, respectively, where the two dipole-allowed transitions $|0\rangle \leftrightarrow |1\rangle$ and $|0\rangle \leftrightarrow |2\rangle$ could be coupled to two metallic waveguides, "port $a$" and "port $b$", with coupling constants $g_a$ and $g_b$, respectively, and the other transition $|1\rangle \leftrightarrow |2\rangle$ with the Rabi frequency $\Omega$ is resonantly driven by a classical driving field with frequency $\nu = \omega_1 - \omega_2$ to connect the two excited states $|1\rangle$ and $|2\rangle$.

The total Hamiltonian of the scheme, which we consider here, in the rotating frame can be expressed as $H = H_W + H_S + H_{W-S}$, where the first term denotes the free Hamiltonian of "port $a$" and "port $b$", which could be metallic waveguides, the second term denotes the Hamiltonian of the V type three-level SQD and the third term is the interaction of SQD and plasmonic waveguides ("port $a$" and "port $b$"), given by as follows ($\hbar = 1$):

$$H_W = i\upsilon_{g,a} \int_{-\infty}^{\infty} dz \left[ a_l^{(a)+}(z)\partial_z a_l^{(a)}(z) - a_r^{(a)+}(z)\partial_z a_r^{(a)}(z) \right]$$
$$+ i\upsilon_{g,b} \int_{-\infty}^{\infty} dz \left[ a_l^{(b)+}(z)\partial_z a_l^{(b)}(z) - a_r^{(b)+}(z)\partial_z a_r^{(b)}(z) \right] \quad (1a)$$

$$H_S = \sum_{i=1,2}(\omega_i - i\Gamma_i/2)|i\rangle\langle i| + \sum_{\substack{i,j=1,2 \\ i\neq j}}\Omega|i\rangle\langle j| \quad (1b)$$

$$H_{W-S} = g_a\left[a_l^{(a)+}(z) + a_r^{(a)+}(z)\right]|0\rangle\langle 1| + g_b\left[a_l^{(b)+}(z) + a_r^{(b)+}(z)\right]|0\rangle\langle 2| + \text{H.c} \quad (1c)$$

, respectively, where $a_r^{(j)+}(z)\left(a_l^{(j)+}(z)\right)(j=a,b)$ is the bosonic creating operator of a right-propagating(left-propagating) surface plasmon in two ports ("port $a$" and "port $b$") at place($z=0$) of the SQD and the surface plasmon with wavevector $k$ propagates with the group velocity, $\upsilon_{g,j}(j=a,b)$ along the interface between the dielectric and metallic medium (i.e., $\omega_k = \upsilon_{g,j}|k|(j=a,b)$). $\omega_i(i=1,2)$ is the excitonic energy corresponding to the state $|i\rangle(i=1,2)$ of the proposed SQD. The total dissipation of each exciton is denoted by the non-Hermitian term $\Gamma_i(i=1,2)$, including the decay rate into other dissipative routes and free space, such as the Ohmic loss. However, it has been familiar that the metallic nanowire could be used to realize the propagation of the surface plasmon even when its radius is small compared to the wavelength of the surface plasmon [11], which describes that the emission from the SQD into surface plasmon modes would be predominate any decay process. Thus, decay rate, $\Gamma_i (i=1,2)$, could be set to be zero in our discussions. $g_j = (2\pi\hbar/\omega_k)^{1/2} \omega_i \mathbf{D}_i \mathbf{e}_k$ denotes the coupling between the SQD and surface plasmon, $\mathbf{D}_i$ denotes the transition dipole moment of the SQD, $\mathbf{e}_k$ denotes the polarization unit vector of propagating the surface plasmon[4].

Supposing that a single plasmon with energy $E_k = \hbar\omega_k$ is injected from the left in "port $a$", then the eigenstate with the single exciton, satisfied $H|\psi_k\rangle = E_k|\psi_k\rangle$, could be taken as follows:

$$|\psi_k\rangle = \int dz [\phi_{k,r}^{(j)+}(z) a_r^{(j)+}(z) + \phi_{k,l}^{(j)+}(z) a_l^{(j)+}(z)]|0,0\rangle + \xi_1|0,1\rangle + \xi_2|0,2\rangle \tag{2}$$

where $|0,0\rangle$ describes the vacuum state without the surface plasmon and the unexcited SQD, $|0,1\rangle(|0,2\rangle)$ describes the vacuum state without the surface plasmon and the SQD being excited in the state $|1\rangle(|2\rangle)$ and $\xi_i (i=1,2)$ denotes the probability amplitude of the excited state $|i\rangle (i=1,2)$ of the SQD. $\phi_{k,r}^{(j)+}(\phi_{k,l}^{(j)+})(j=a,b)$ denotes the wavefunction of a right-propagating (a left-propagating) single plasmon in the plasmonic waveguide, "port $j$" $(j=a,b)$ at place z.

For a single plasmon injected from the left in "port $a$", the wavefunctions $\phi_{k,r}^{(j)+}(z)$ and $\phi_{k,l}^{(j)+}(z)$ could be taken as follows:

$$\phi_{k,r}^{(a)+}(z) = \begin{cases} e^{ikz}, & z<0, \\ t^{(a)} e^{ikz} & z>0, \end{cases} \quad \phi_{k,l}^{(a)+}(z) = \begin{cases} r^{(a)} e^{-ikz}, & z<0, \\ 0 & z>0, \end{cases} \tag{3}$$

$$\phi_{k,r}^{(b)+}(z) = \begin{cases} 0 & z<0, \\ t_r^{(b)} e^{ikz} & z>0, \end{cases} \quad \phi_{k,l}^{(b)+}(z) = \begin{cases} 0, & z<0, \\ t_l^{(b)} e^{-ikz} & z>0, \end{cases} \tag{4}$$

where $t^{(a)}$ ($r^{(a)}$) denotes the transmission (reflection) amplitude in "port $a$" at the place $z$ and $t_r^{(b)}(t_l^{(b)})$ is the forward (backward) transfer amplitude in "port $b$". By substituting Eq.(2) into $H|\psi_k\rangle = E_k|\psi_k\rangle$ and applying mode functions with continuity conditions $t^{(a)} = 1 + r^{(a)}$ and $t_l^{(b)} = t_r^{(b)} = t^{(b)}$, we obtain the reflection amplitude in "port $a$" and the transfer amplitude in "port $b$", respectively, as

$$r^{(a)} = -\frac{J_a(J_b + i\delta_2)}{(J_a + i\delta_1)(J_b + i\delta_2) + \Omega^2}, \tag{5}$$

$$t^{(b)} = \frac{ig_a J_b \Omega}{g_b(J_a + i\delta_1)(J_b + i\delta_2) + \Omega^2}, \tag{6}$$

where $J_j \equiv g_j^2/\upsilon_{g,j}$ $(j=a,b)$ and $\delta_i = \omega_k - \omega_i$ $(i=1,2)$. As we can see easily from Eqs. (5) and (6), single plasmons injected from "port $a$" could be reflected or transmitted only under the condition without classical field, but the effect of the classical field makes the injected plasmon in "port $a$" transfer to "port $b$", which means that it is just the classical field to change the propagation direction of the single plasmon in the quantum router.

## 3. Numerical Analysis and Theoretical Results

The routing properties of the proposed system for the injected single plasmons in a long time limit is characterized by the transmission coefficient $T^{(a)} \equiv |t^{(a)}|^2$, the reflection coefficient $R^{(a)} \equiv |r^{(a)}|^2$ and the transfer rate $T^{(b)} \equiv |t^{(b)}|^2$.

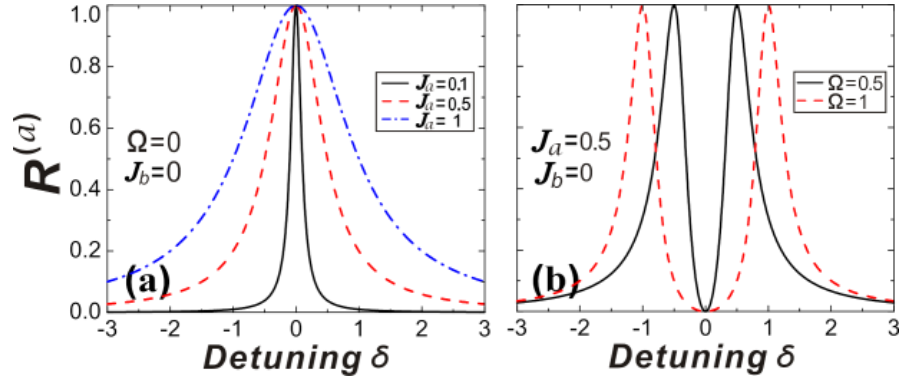

**Fig. 2** (color online). Reflection spectrum of single plasmons propagating along the metallic waveguide, "port $a$", interacting with the three-level SQD as function of detuning $\delta$. (a) $J_a$=0.1(black solid line), 0.5(red dashed line), 1(blue dash-dotted line) and $\Omega$ =0, $J_b$=0; (b) $\Omega$ =0.5 (black solid line), 1(red dashed line) and $J_a$ =0.5, $J_b$=0. In all cases, we set $\delta_1= \delta_2= \delta$ and the parameters such as the detunings, the intensity of classical field and couplings are in units of $10^{-4}\, \omega_0$, where $\omega_0=2\pi\upsilon_g/L$.

In Fig. 2, we have a graphical illustration of the reflection spectrum of single plasmons injected from the metallic waveguide, "port $a$" interacting with a three-level SQD as function of detuning with turning on and off the classical field and no coupling between the SQD and the metallic waveguide, "port $b$", where we set $\delta_1= \delta_2= \delta$. Fig. 2(a) shows the reflection spectrum of the injected single plasmons in "port $a$" with turning off the classica l driving field (i.e., $\Omega = 0$), which is corresponding to the transport of the single plasmon interacting with two level SQD coupled to one plasmonic waveguide. As in Fig. 2(a), the width of reflection spectrum gets wider as the coupling $J_a$ between the SQD and "port a" becomes stronger and there appears only one completed reflection peak at $\delta = 0$, which is the same result as in Ref. [11,13,14], as we expected. This results is held even when the coupling $J_b$ between the SQD and "port b" is not zero, which means the single plasmons cannot be transferred from "port a" to "port b". In Fig. 2(b) we can find the reflection spectrum of the injected single plasmons in "port a" with turning on the classic driving field, there exists the completed transmission dip in resonance and two completed reflection peaks are occurred at $\delta = \pm\Omega$, which is called Aulter-Townes splitting. As you can see from Fig. 2(b), of course, there is no transfer from "port a" to "port b" and there appears the Rabi oscillation only between two excited states. From Fig. 2 we can find that single plasmons cannot be transferred from "port a" to "port b" in cases of turning off the classical driving field or no coupling between the SQD and the metallic waveguide, "port b", resulting in the transport conservation equation $T^{(a)} + R^{(a)} = 1$ and that the switching between the completed transmission and reflection could be realized and could be controllable the separation of completed reflection peaks by adjusting the classic driving field.

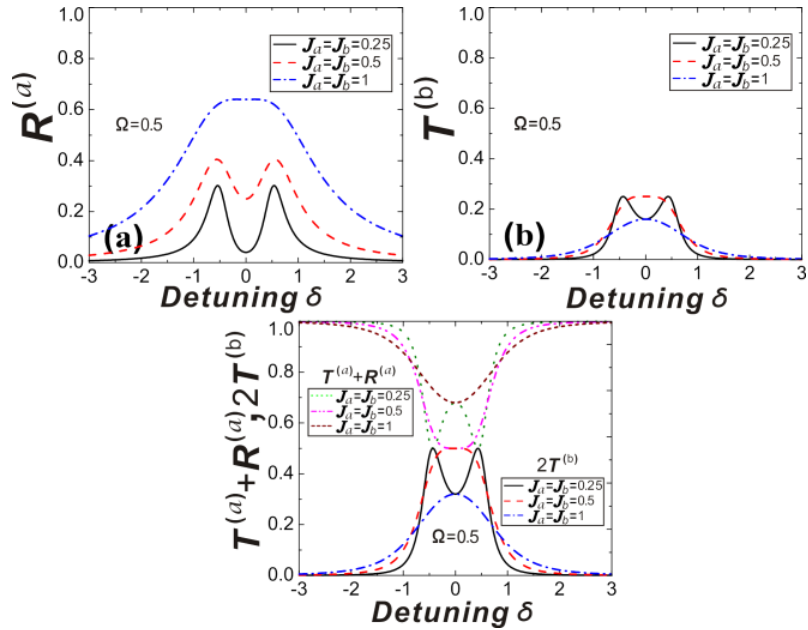

**Fig. 3** (color online). (a) Reflection spectra of propagating single plasmon in "port a" for different couplings between the SQD and two ports, where $J_a=J_b=0.25$(black solid line), 0.5(red dashed line), 1(blue dash-dotted line) versus detuning $\delta$. (b) Transfer rate of single plasmon in one of two direction of "port *b*" for different couplings between the SQD and two ports, where $J_a=J_b$=0.25(black solid line), 0.5(red dashed line), 1(blue dash-dotted line) versus detuning $\delta$. (c) The transport rate of single plasmons in "port *a*" and "port *b*" for different couplings between the SQD and two ports, where $J_a=J_b$=0.25(green dotted line and black solid line), 0.5(pink dash-dot-dotted line and red dashed line), 1(wine short dashed line and blue dash-dotted line) versus detuning $\delta$. In all cases, we set $\delta_1=\delta_2=\delta$ and $\Omega$ =0.5. In all the cases, the parameters such as the detunings, the intensity of classical field and couplings are in units of $10^{-4}$ $\omega_0$, where $\omega_0=2\pi\upsilon_g/L$.

When single plasmons come from the left of one plasmonic waveguide (e.g., "port *a*"), it will be absorbed by the SQD, which is then excited from the ground state $|0\rangle$ to the excited state $|1\rangle$. Since the transition between excited states of the SQD, $|1\rangle$ and $|2\rangle$ is resonantly driven by the classic driving field, the excited SQD will be coupled to either "port *a*" or "port *b*". Thus, mediated by the SQD with V type three-level, single plasmons could be guided from one plasmonic waveguide to the other. In other words, the resonant tunneling transition of the excited states helps the SQD to be exploited to redirect single plasmons propagating from one quantum channel to the other. To investigate this process, we plot the current flow of single plasmons in "port *a*" and "port *b*" in Fig. 3, where the couplings between the SQD and two ports ("port *a*" and "port *b*") are supposed to be the same each other, resulting in the transport conservation equation $T^{(a)}+ R^{(a)}+2T^{(b)} =1$. From Fig. 3(a) we can see there exist two reflection peaks of single plasmons in "port *a*" when coupling between the SQD and two ports is 0.25(black solid line) and 0.5(red dashed line), respectively, and it vanishes when the coupling becomes stronger and there appears only one broadened reflection peak with the coupling strength of 1. The reflection spectrum gets higher as the coupling between the SQD and two ports becomes stronger and there appears only one completed reflection peak. Fig. 3(b) shows the transfer rate of the single plasmon from "port *a*" to "port *b*" as function of detuning, where there exist two peaks of the transfer rate until the coupling between the SQD and two ports approaches to 0.5 and it has the highest broadened peak when the value is consisted with the strength of the classic driving field. When the coupling between the SQD and two ports is over 0.5, the peak of the transfer rate gets down. Fig. 3(c) is plotted the total scattering spectra in "port *a*" and "port *b*", respectively, where it is found that when the coupling $J_a=J_b$ goes to the intensity of the classical driving field, the maximum of transfer rate of the splitted doublet is equal t

o 0.5 and its separation is fixed, which means the half of incoming single plasmons could be transferred from "port $a$" to "port $b$". We notice that one could be controllable the width of the transfer peak (red dashed line) by adjusting several parameters such as the coupling between the SQD and two ports, the intensity of the classical field, $\Omega$, which implies one could be achieved the routing of single plasmons not only in a specific frequencies but also in a wide-band frequency region. Generally, an optical pulse could be actually a superposition of plan waves with different frequencies where the off-resonant components can deviate from the expected transfer rate dramatically. Therefore, the wide-band transfer spectrum obtained could be used to the practical possibilities for realizing a single plasmon router.

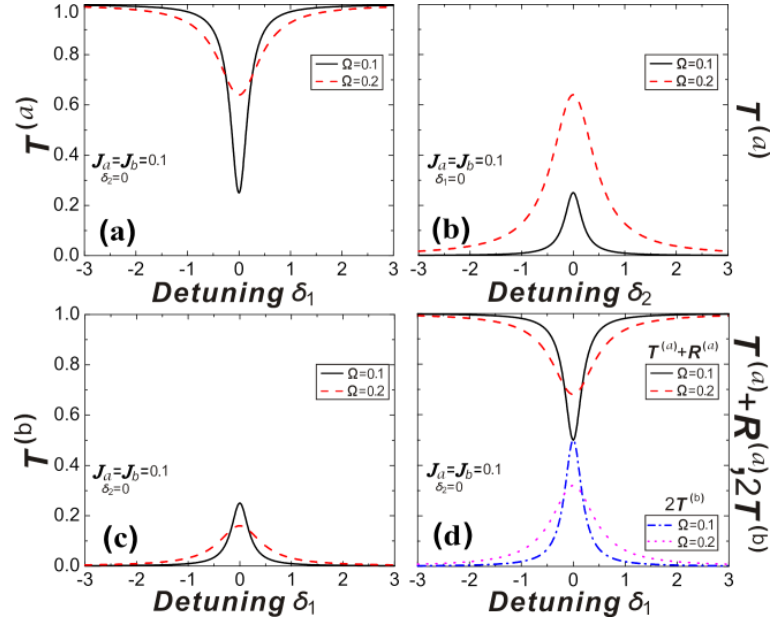

**Fig. 4** (color online). Transmission spectra of propagating single plasmon in "port $a$" (a) according to detuning $\delta_1$, $\delta_2=0$, (b) according to detuning $\delta_2$, $\delta_1=0$, for classic field with $\Omega=0.1$(black solid line), 0.2(red dashed line). (c) Transfer rate of single plasmon in one of two direction of "port $b$" for classic field with $\Omega=0.1$(black solid line), 0.2(red dashed line) versus detuning $\delta_1$ and $\delta_2=0$. (d) The transport rate of single plasmons in "port $a$" and "port $b$" for classic field with $\Omega=0.1$(blue dashed line and black solid line), 0.2(pink dotted line and red dashed line) versus detuning $\delta_1$ and $\delta_2=0$. In all cases, we set $J_a=J_b=0.1$. In all the cases, the parameters such as the detunings, the intensity of classical field and couplings are in units of $10^{-4}\omega_0$, where $\omega_0=2\pi v_g/L$.

Fig. 4 shows the routing functions of our proposed system for the injected single plasmons from the left of "port $a$" when one of the transitions of the SQD is coupled to one of ports with the variable detuning and the other transition of that is resonantly coupled to the other port. In Fig. 4(a), 4(c) and 4(d), we plot the transmission coefficient $T^{(a)}$ in "port $a$", transfer rate $T^{(b)}$ from "port $a$" to "port $b$"and total scattering spectra in "port $a$" and "port $b$", respectively, with different intensities of classical field, $\Omega$ when the transitions $|0\rangle \leftrightarrow |1\rangle$ of the SQD is coupled to "port $a$" with detuning and the other transition $|0\rangle \leftrightarrow |2\rangle$ of the SQD is resonantly coupled to "port $b$". On the contrary, Fig. 4(b) shows the transmission coefficient $T^{(a)}$ in "port $a$" as a function of detuning $\delta_2$ with different intensities of classical field, $\Omega$ when the transitions $|0\rangle \leftrightarrow |2\rangle$ of the SQD is non-resonantly coupled to "port $b$" with detuning and the other transition $|0\rangle \leftrightarrow |1\rangle$ of the SQD is coupled to "port $a$" with no detuning. In Fig. (4), all the curves have symmetric shape for the resonant energy with the injected single plasmons. As shown in Fig. 4(a) and 4(b), we can find there appears one dip or peak in the transmission spectra of single plasmons in "port $a$" according to which transition of the SQD is resonant with the injected single plasmons, where the dip gets shallower or the peak gets higher when the classic driving field becomes strong. As the same as results in Fig. 3(c), the transfer rate of single plasmons from "port $a$" to "port $b$" has maximum values when the coupling between the SQD and two ports is consisted with the classic driving field.

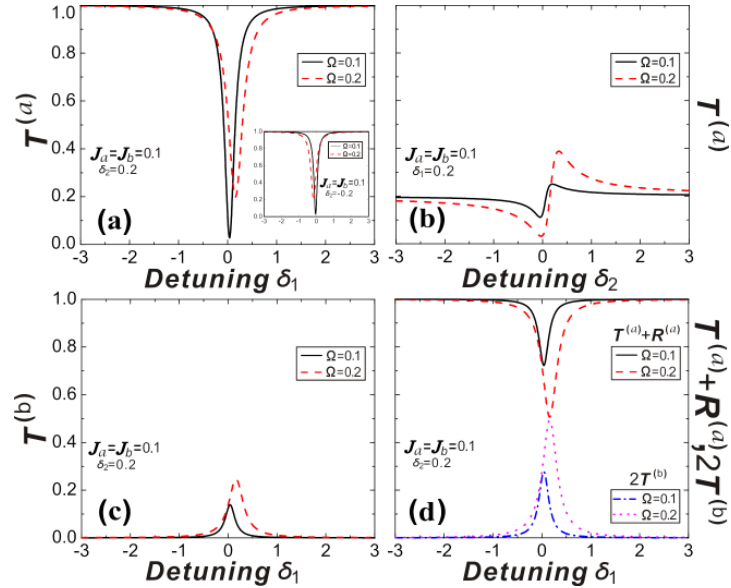

**Fig. 5** (color online). Transmission spectra of propagating single plasmon in "port $a$" (a) according to detuning $\delta_1$, $\delta_2$=0.2, inset with $\delta_2$=-0.2, (b) according to detuning $\delta_2$, $\delta_1$=0.2, for classic field with $\Omega$ =0.1(black solid line), 0.2(red dashed line). (c) Transfer rate of single plasmon in one of two direction of "port *b*" for classic field with $\Omega$ =0.1(black solid line), 0.2 (red dashed line) versus detuning $\delta_1$ and $\delta_2$=0.2. (d) The transport rate of single plasmons in "port *a*" and "port *b*" for classic field with $\Omega$ =0.1(blue dashed line and black solid line), 0.2(pink dotted line and red dashed line) versus detuning $\delta_1$ and $\delta_2$=0.2. In all cases, we set $J_a$=$J_b$=0.1. In all the cases, the parameters such as the detunings, the intensity of classical field and couplings are in units of $10^{-4}$ $\omega_0$, where $\omega_0$=2π$\upsilon_g$/L.

Finally, in Fig. 5 we investigate the transmission $T^{(a)}$ in "port *a*", transfer rate $T^{(b)}$ from "port *a*" to "port *b*", total transport in "port *a*" and "port *b*", respectively, with different intensities of classical field, $\Omega$ when one of the transitions of the SQD is coupled to one of ports with the variable detuning and the other transition of that is coupled to the other port with a certain detuning. As you can see from Fig. 5(a) with the inset, 5(c) and 5(d), there appears one dip or one peak in spectra of $T^{(a)}$, $T^{(b)}$, $T^{(a)}$+ $R^{(a)}$, 2$T^{(b)}$ with a certain classic driving field, where it move to the left or right form the resonant energy when the transitions $|0\rangle \leftrightarrow |2\rangle$ of the SQD is coupled to "port *a*" with red shift or blue one. In comparison with the above result, the transfer rate of single plasmons from "port *a*" to "port *b*" has maximum values when the coupling between the SQD and two ports is different with the classic driving field in Fig. 5(d). We can find the scattering property of single plasmons in "port *a*" is greatly changed when the detuning between the transitions $|0\rangle \leftrightarrow |2\rangle$ of the SQD and the energy of the injected single plasmons is variant and the other transition $|0\rangle \leftrightarrow |1\rangle$ of the SQD is coupled to "port *a*" with a certain detuning, where the transmission spectra have Fano-line shape, which couldn't find in previous investigations[25-27,30]. Anyway, from the above results we found that the transfer rates form "port *a*" to "port *b*" and the peak positions of the transfer rates could be controlled by adjusting the parameters such as the detuning between the SQD and two port, and the intensity of the classical field, $\Omega$.

## 4. Conclusions

In summary, we investigated theoretically the transport properties of single plasmons interacting the single self-assembled InGaAs/GaAs semiconductor quantum dot with a V type three-level sandwiched between two metallic waveguides. On the basis of such a hybrid sysem, we find single plasmons from the input port could be switchable and redirected by controlling parameters, such as the intensity of the classical field, the detunings, and the interaction between the SQD and the waveguides. In particular, single plasmons propagating in a port could be transferred to the other one by applying the classical field to the SQD, which suggests the classical field could be used to achieve the quantum routing function. We also note that suitable setting the parameters could result in a single plasmon router with a wide-band transfer spectrum, which is valuble for practical applications. Our proposed scheme for a single plasmon router could find further potential possibilities for realizing nano-optical and quantum information processing devices, such as quantum switches, quantum logic gates, directional coupler.

**Acknowledgments.** This work was supported by the National Program on Key Science Research of DPR of Korea (Grant No. 1-6-8).